\documentclass[aps,prl,reprint,showpacs]{revtex4-1}

\usepackage{graphicx}
\usepackage{subfigure}
\usepackage{dcolumn}
\usepackage{verbatim}%for multiple line commenting
\usepackage{amsmath}
\usepackage{amssymb}
\usepackage{float}% floating figures at the exact position
\usepackage[colorlinks=true,linkcolor=blue,citecolor=blue,filecolor=blue,urlcolor=blue]{hyperref}
\begin{document}

\title{A Model for Large $\theta_{13}$ Constructed using the Eigenvectors of the $S_4$ Rotation Matrices}

\author{R.~Krishnan}
\email[]{k.rama@warwick.ac.uk}
\affiliation{Department of Physics, University of Warwick, Coventry CV4 
7AL, UK}

\date{\today}

\begin{abstract}
A procedure for using the eigenvectors of the elements of the representations of a discrete group in model building is introduced and is used to construct a model that produces a large reactor mixing angle, $\sin^2\theta_{13}=\frac{2}{3}\sin^2\frac{\pi}{16}$, in agreement with recent neutrino oscillation observations. The model fully constrains the neutrino mass ratios and predicts normal hierarchy with the light neutrino mass, $m_1\approx 25~\text{meV}$. Motivated by the model, a new mixing ansatz is postulated which predicts all the mixing angles within $1\sigma$ errors.
\end{abstract}

\maketitle

\begin{figure}[H]
\begin{center}
\includegraphics[scale=1.0]{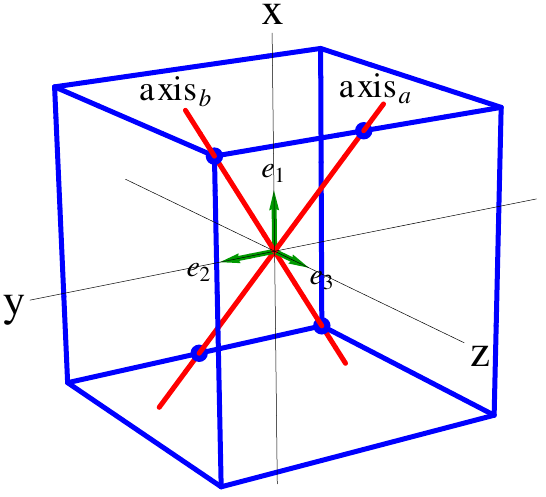}
\caption{The generators $a$ and $b$ represent $\pi$-rotation about $\text{axis}_a$ and $\frac{2\pi}{3}$-rotation about $\text{axis}_b$ respectively.}
\label{fig:cube}
\end{center}
\end{figure}

\noindent{\bf Introduction}: We use the group $SU(3)$ and its discrete subgroup $S_4$ for model building. Both of these groups had been studied extensively as flavour symmetry groups, e.g.~\cite{[{$SU(3)$: }]SU31,*SU32,*SU33,*SU34,*SU35}~\cite{[{$S_4$: }]S41,*S42,*S43,*S44,*S45}. The $S_4$ group has the presentation \cite{book}
\begin{equation}
\langle a,b | a^2 = b^3 = (ab)^4 = e \rangle.
\end{equation}
$S_4$ is the symmetry group of the cube (Fig.~\ref{fig:cube}) and the elements of the group can be represented as the orientation preserving rotations of the cube. The matrices representing the generators can be written as
\begin{equation}\label{eq:gen}
a =
\left(\begin{matrix}0 & 0 & 1\\
       0 & \text{-}1 & 0\\
       1 & 0 & 0
\end{matrix}\right)\\, \quad b=
\left(\begin{matrix}0 & 0 & 1\\
       1 & 0 & 0\\
       0 & 1 & 0
\end{matrix}\right).
\end{equation}
Here the basis vectors $e_1=(1,0,0)^T$, $e_2=(0,1,0)^T$ and $e_3=(0,0,1)^T$ form the symmetry axes of the cube passing through face centres. If we define the left-handed leptons, $L=(L_e, L_\mu, L_\tau)^T$ where $L_e=(\nu_{eL},e_L)^T$ etc., as a triplet in this basis, the flavour states $L_e$, $L_\mu$, $L_\tau$ correspond to $e_1$, $e_2$, $e_3$ respectively. Usually in models a set of flavons are introduced whose vacuum expectation values (vevs) produce the desired texture for the fermion mass matrices. In other words, the orientation of fermion flavour states as well as the flavon vevs in the flavour space determines the form of the mass matrices.

Axes of the orientation preserving rotations of the cube are nothing but eigenvectors of the corresponding rotation matrices with eigenvalue equal to +1. The basis vectors $e_1$, $e_2$ and $e_3$ are examples. There are also other vectors like the ones passing through the opposite edge centres (e.g.~$\text{axis}_a$ in Fig.~\ref{fig:cube}) and the ones passing through the opposite vertices (e.g.~$\text{axis}_b$ in Fig.~\ref{fig:cube}). Compared to vectors pointing in random directions, these vectors are ``special" in the context of the $S_4$ symmetry. The rotation matrices are unitary and so their eigenvalues in general are complex numbers with unit modulus. If non-degenerate, these eigenvalues also correspond to unique eigenvectors and the author argues that these eigenvectors are also ``special'' like the rotation axes.

As an example consider $v=\frac{1}{\sqrt{3}}(1,\bar{\omega},\omega)^T$, the normalised ($v^\dagger v=1$) eigenvector of the matrix $b$ in Eq.~(\ref{eq:gen}), corresponding to the eigenvalue $\omega$, where $\omega=e^{i\frac{2\pi}{3}}$ and $\bar{\omega}=e^{\text{-}i\frac{2\pi}{3}}$. Since $e^{i\theta} v$, where $e^{i\theta}$ is an arbitrary phase, is also a normalised eigenvector, we impose the following condition to uniquely fix the phase: The component of the eigenvector in the direction of one of the basis vectors should have zero phase ie.~$\text{arg}(v^\dagger e_i)=0$, where $i=1$, $2$ or $3$. This is intuitive since the basis vectors are used to define the fermion flavour states. Imposing the above mentioned phase condition, the allowed choices for $v$ are $\frac{1}{\sqrt{3}}(1,\bar{\omega},\omega)^T$, $\frac{1}{\sqrt{3}}(\omega,1,\bar{\omega})^T$ and $\frac{1}{\sqrt{3}}(\bar{\omega},\omega,1)^T$.

Let $g$ represent an element of the group and $l$ be one of its non-degenerate eigenvalues. The corresponding ``special'' eigenvector, $\text{eig}(g,l)_i$, is defined using the following normalisation condition and the phase condition:
\begin{equation}
\text{eig}(g,l)_i^\dagger\,\text{eig}(g,l)_i=1,\quad \text{arg}\left(\text{eig}(g,l)_i^\dagger \, e_i\right)=0.
\end{equation}
For example, the basis vector $e_3$ is $\text{eig}(ab,1)_3$ where
\begin{equation}
ab =
\left(\begin{matrix}0 & 1 & 0\\
       \text{-}1 & 0 & 0\\
       0 & 0 & 1
\end{matrix}\right)
\end{equation}
represents $\frac{\pi}{2}$-rotation about Z-axis. We define the group elements
\begin{equation}\label{eq:gen}
c=bab =
\left(\begin{matrix}0 & 0 & 1\\
       0 & 1 & 0\\
       \text{-}1 & 0 & 0
\end{matrix}\right)\\, \quad d=a(bab)^2=
\left(\begin{matrix}0 & 0 & \text{-}1\\
       0 & \text{-}1 & 0\\
       \text{-}1 & 0 & 0
\end{matrix}\right).
\end{equation}
The matrix $c$ represents $\frac{\pi}{2}$-rotation about Y-axis. The matrix $d$ represents $\pi$-rotation about the axis which lies in the X-Z plane and which is perpendicular to $\text{axis}_a$, ie.~just like $\text{axis}_a$, this axis also passes through a set of opposite edge centres of the cube. The eigenvectors of $a$, $b$, $c$ and $d$ which will be later used in model building are listed below:
\begin{equation}
\begin{aligned}
\text{eig}(b,1)_1&=\tfrac{1}{\sqrt{3}}\left(1,1,1\right)^T, &\text{eig}(a,1)_1&=\tfrac{1}{\sqrt{2}}\left(1,0,1\right)^T, \\
\text{eig}(b,\omega)_1&=\tfrac{1}{\sqrt{3}}\left(1,\bar{\omega},\omega\right)^T, &\text{eig}(c,i)_1&=\tfrac{1}{\sqrt{2}}\left(1,0,i\right)^T, \\
\text{eig}(b,\bar{\omega})_1&=\tfrac{1}{\sqrt{3}}\left(1,\omega,\bar{\omega}\right)^T, &\text{eig}(c,i)_3&=\tfrac{1}{\sqrt{2}}\left(\text{-}i,0,1\right)^T,\\
\text{eig}(d,1)_1&=\tfrac{1}{\sqrt{2}}\left(1,0,\text{-}1\right)^T, &\text{eig}(c,\text{-}i)_1&=\tfrac{1}{\sqrt{2}}\left(1,0,\text{-}i\right)^T, \\
\text{eig}(d,1)_3&=\tfrac{1}{\sqrt{2}}\left(\text{-}1,0,1\right)^T, &\text{eig}(c,\text{-}i)_3&=\tfrac{1}{\sqrt{2}}\left(i,0,1\right)^T.
\end{aligned}
\label{eq:eiglist}
\end{equation}

\noindent{\bf The Model}: Recent experiments \cite{DayaBay,*RENO,*DCHOOZ,*T2K,*MINOS} have shown that the reactor mixing angle, $\theta_{13}$, is non-zero. A number of models and parameterisations have been proposed, e.g.~\cite{M0,*M1,*M2,*M3,*M4,*M5,*M6}, to accommodate the non-zero $\theta_{13}$. The model described in this paper is constructed in the Standard Model (SM) framework with the addition of the right-handed neutrino triplet, $\nu_R=(\nu_{1R}, \nu_{2R}, \nu_{3R})^T$ in the context of the type-1 seesaw mechanism. We postulate a global flavour symmetry group, 
\begin{equation}\label{eq:Gf}
G_f=SU(3)_1\times SU(3)_2\times U(1)_f. 
\end{equation}
The fermion fields and a set of postulated flavons belong to specific representations of $G_f$ as shown in Table~\ref{tab:reps}. The $U(1)$ group is introduced to ensure that the flavons couple to only the desired fermions. We write the mass terms at the lowest order, containing the fermions and the minimum number of flavons, invariant under $G_f$ and the SM gauge group. The eigenvectors of the elements of the $S_4$ subgroup of the $SU(3)$ group are used to construct the vevs of the flavons. These vevs produce the required mass matrices.

{\renewcommand{\arraystretch}{1.2}
\begin{table}[H]
\begin{center}
\begin{tabular}{| c | c | c | c |c | c | c | c | c | c | c| c |}
\hline
		& $L$ 		& $e_R$	& $\mu_R$	& $\tau_R$	&$\nu_R$ 	&$\phi_e$ 	&$\phi_\mu$	&$\phi_\tau$ 	&$\phi$ 	&$\xi_1$	& $\xi_2$\\
\hline
$SU(3)_1$ 	& $\boldsymbol{3}$ & $\boldsymbol{1}$ & $\boldsymbol{1}$ & $\boldsymbol{1}$ &$\boldsymbol{\bar{3}}$ &$\boldsymbol{3}$ &$\boldsymbol{3}$ &$\boldsymbol{3}$ &$\boldsymbol{3}$ & $\boldsymbol{6}$ &$\boldsymbol{1}$\\
\hline
$SU(3)_2$ 	& $\boldsymbol{1}$ & $\boldsymbol{1}$ & $\boldsymbol{1}$ & $\boldsymbol{1}$ &$\boldsymbol{1}$ &$\boldsymbol{1}$ &$\boldsymbol{1}$ &$\boldsymbol{1}$ &$\boldsymbol{3}$ & $\boldsymbol{1}$ &$\boldsymbol{\bar{6}}$\\
\hline
$U(1)_f$ 	& $f_l$ & $f_l+f_e$ & $f_l+f_\mu$ & $f_l+f_\tau$ &$0$ &$-f_e$ &$-f_\mu$ &$-f_\tau$ &$0$ & $f_l$ &$0$\\
\hline
\end{tabular}
\end{center}
\caption{The fields $\phi_e$, $\phi_\mu$, $\phi_\tau$, $\phi$, $\xi_1$, $\xi_2$ are the flavons. For the $U(1)$ group the tabulated values are the generators, e.g.~$f_l\equiv e^{i f_l \theta}$ . The SM Higgs is a flavour singlet.}
\label{tab:reps}
\end{table}}

The tensor product expansion of the fundamental representations of $SU(3)$ are
\begin{equation}
\boldsymbol{3}\otimes\boldsymbol{3}=\boldsymbol{\bar{3}}\oplus\boldsymbol{6},\quad\boldsymbol{\bar{3}}\otimes\boldsymbol{\bar{3}}=\boldsymbol{3}\oplus\boldsymbol{\bar{6}},\quad\boldsymbol{3}\otimes\boldsymbol{\bar{3}}=\boldsymbol{1}\oplus\boldsymbol{8}.
\end{equation}
For the charged leptons, the lowest order mass term is
\begin{equation}\label{eq:mclept}
\overline{L} \left(y_e \phi_e e_R +y_\mu \phi_\mu \mu_R +y_\tau \phi_\tau \tau_R \right)H\tfrac{1}{\Lambda_{\phi_l}} + H.C.
\end{equation}
where $y_\alpha$ are the coupling constants, $H$ is the SM Higgs, $\Lambda_{\phi_l}$ is the cut-off scale for the flavons $\phi_e$, $\phi_\mu$ and $\phi_\tau$. The $S_4$ group has four irreducible representations: $\boldsymbol{1}$, $\boldsymbol{2}$, $\boldsymbol{3}$ and $\boldsymbol{3'}$ \cite{S4Table}. The orientation preserving rotations of the cube discussed earlier belong to $\boldsymbol{3}'$. The flavon triplets $\phi_e$, $\phi_\mu$ and $\phi_\tau$ belong to the representation $\boldsymbol{3}$ of $SU(3)_1$. The restriction of the representation $\boldsymbol{3}$ of $SU(3)$ to its subgroup $S_4$ is the representation $\boldsymbol{3'}$ of $S_4$. Therefore the ``special'' eigenvectors of the representation matrices of $\boldsymbol{3'}$ of $S_4$ are used to construct the vevs of the flavons. We assign:
\begin{equation}
\langle\phi_e\rangle= \text{eig}(b,1)_1, \quad \langle\phi_\mu\rangle=\text{eig}(b,\omega)_1, \quad \langle\phi_\tau\rangle=\text{eig}(b,\bar{\omega})_1
\end{equation}
where the angular brackets are used to denote vevs. We do not discuss the mechanism of flavour symmetry breaking in this paper. To avoid Goldstone bosons, it is necessary to add explicit symmetry breaking terms for the flavon potentials, which break the continuous flavour group $G_f$, Eq.~(\ref{eq:Gf}), into an unknown discrete group. The flavon vevs spontaneously break this discrete flavour symmetry. Also the Higgs vev, $(0,h_o)^T$, breaks the weak gauge symmetry. After the symmetry breaking, the charged-lepton mass term, Eq.~(\ref{eq:mclept}), takes the form
\begin{equation}
\overline{l_L} \mathcal{T}^\dagger M_d l_R + H.C
\end{equation}
where  
\begin{equation}
\mathcal{T}^\dagger=\frac{1}{\sqrt{3}}\left(\begin{matrix}1 & 1 & 1\\
1 & \bar{\omega} & \omega\\
1 & \omega & \bar{\omega}
\end{matrix}\right),
\end{equation}
$l_L=(e_L, \mu_L, \tau_L)^T$, $l_R=(e_R, \mu_R, \tau_R)^T$ and $M_d=\text{diag}(m_e,m_\mu,m_\tau)$ with $m_e=\frac{y_e h_o}{\Lambda_{\phi_e}}$ etc. The charged-lepton mass matrix, $\mathcal{T}^\dagger M_d$, when left-multiplied with $\mathcal{T}$, is diagonalised giving the charged-lepton masses $m_e$, $m_\mu$, $m_\tau$. $\mathcal{T}$ is the Trimaximal mixing matrix \cite{TM}. 

For the neutrinos, the lowest order Dirac mass term is
\begin{equation}\label{eq:mdl2dirac}
2y_w \overline{L} \xi_1 \nu_R \tilde{H} \tfrac{1}{\Lambda_{\xi_1}} +H.C.
\end{equation}
where $\tilde{H}$ is the conjugate Higgs, $y_w$ is the coupling and $\Lambda_{\xi_1}$ is the cut-off scale for the flavon $\xi_1$. The flavon $\xi_1$ belongs to the representation $\boldsymbol{6}$ of $SU(3)_1$ and can be written as a $3\times3$ complex-symmetric matrix. The restriction of $\boldsymbol{6}$ of $SU(3)$ to the $S_4$ subgroup is the direct sum of $\boldsymbol{1}$, $\boldsymbol{2}$ and $\boldsymbol{3}$ of $S_4$. We assign a very simple choice of vev for $\xi_1$ where $\boldsymbol{2}$ and $\boldsymbol{3}$ parts vanish, ie.~$\langle\xi_1\rangle$ becomes the identity when written in the matrix form. After the symmetry breaking, the Dirac mass term, Eq.~(\ref{eq:mdl2dirac}), takes the form
\begin{equation}
m_w \left(\overline{\nu_L} \nu_R + \overline{(\nu_R)^c} (\nu_L)^c \right)+ H.C.
\label{eq:dirac}
\end{equation}
where $\nu_L=(\nu_{eL}, \nu_{\mu L}, \nu_{\tau L})^T$ and $m_w=\frac{y_w h_o}{\Lambda_{\xi_1}}$. 

The lowest order Majorana mass term for the neutrinos is
\begin{equation}\label{eq:majterm}
y_G \overline{(\nu_R)^c} \phi \xi_2 \phi^T \nu_R \tfrac{1}{\Lambda_{\xi_2} \Lambda_\phi^2}+ H.C.
\end{equation}
where $\Lambda_{\xi_2}$ and $\Lambda_\phi$ are the cut-off scales for the flavons $\xi_2$ and $\phi$ respectively. Note that $\phi$ transforms as a $\boldsymbol{3}$ under both $SU(3)_1$ and $SU(3)_2$. Therefore $\phi$ can be written as a $3\times3$ matrix, $\phi_{ij}$, the row index $i$ representing $SU(3)_1$ and the column index $j$ representing $SU(3)_2$. The flavon $\xi_2$ belongs to the representation $\boldsymbol{\bar{6}}$ of $SU(3)_2$. A $\boldsymbol{\bar{6}}$ of $SU(3)$, just like a $\boldsymbol{6}$, contains a $\boldsymbol{1}$, a $\boldsymbol{2}$ and a $\boldsymbol{3}$ of the $S_4$ subgroup. As was done earlier for the case of $\langle\xi_1\rangle$, here we assign $\langle\xi_2\rangle$ also to be equal to the identity. After the symmetry breaking, the Majorana mass term, Eq.~(\ref{eq:majterm}), takes the form
\begin{equation}
m_G \overline{(\nu_R)^c} \langle\phi\rangle\langle\phi\rangle^T \nu_R+ H.C.
\label{eq:majorana}
\end{equation}
where $m_G = \frac{y_G}{\Lambda_{\xi_2} \Lambda_\phi^2}$. The matrix $\langle\phi\rangle\langle\phi\rangle^T$ is complex-symmetric and $\langle\phi\rangle\langle\phi\rangle^T$ contains all the interesting physics in our model. %Four related choices of vevs for $\phi$ are studied in the following discussion. 

%Note that the flavon $\phi$ is a $\boldsymbol{3}$ under both $SU(3)_1$ and $SU(3)_2$. So $\phi$ can be written as a $3\times3$ matrix, $\phi_{ij}$, the row and the column indices corresponding to $SU(3)_1$ and $SU(3)_2$ respectively. 

To assign vev for the flavon $\phi_{ij}$, we use $(S_4)_1\times (S_4)_2$, the subgroup of $SU(3)_1\times SU(3)_2$. The group $(S_4)_1\times (S_4)_2$ has $24\times24=576$ elements. Let $g_1$ and $g_2$ be the elements of $(S_4)_1$ and $(S_4)_2$ respectively. If $v_1$ and $v_2$ are the eigenvectors of $g_1$ and $g_2$ corresponding to the eigenvalues $a_1$ and $a_2$, then the direct product $v_1\times v_2$ will be an eigenvector of $g_1\times g_2$ with an eigenvalue $a_1 a_2$. Now we make the following assumption:
\begin{equation}
\langle \phi \rangle = v^1_1\times v^1_2~+~v^2_1\times v^2_2~+~v^3_1\times v^3_2~+~v^4_1\times v^4_2
\label{eq:sumvev}
\end{equation}
where the RHS of Eq.~(\ref{eq:sumvev}) is the sum of four eigenvectors. Based on the choices for $v^i_1\times v^i_2$s we get a set of similar cases of solutions described in the following sections. The assumed form of $\langle \phi \rangle$ given in Eq.~(\ref{eq:sumvev}) and the choices for $v^i_1\times v^i_2$s were obtained through educated guesses and also through trial and error to fit the experimental data.

\noindent {\it \bf Case 1}: Here we assign
\begin{equation}
\begin{aligned}
v^1_1&=e_2, &v^1_2&=e_2,\\
v^2_1&=\text{eig}(a,1)_1, &v^2_2&=e_1,\\
v^3_1&=\text{eig}(d,1)_1, &v^3_2&=e_1,\\
v^4_1&=\text{eig}(d,1)_3, &v^4_2&=\text{eig}(c,\text{-}i)_1.
\end{aligned}
\label{eq:case1vev}
\end{equation}
Using Eq.~(\ref{eq:sumvev}), Eqs.~(\ref{eq:case1vev}) and Eqs.~(\ref{eq:eiglist}), we get~\footnote{The set of eigenvectors given in Eqs.~(\ref{eq:case1vev}) is not the only choice that results in the matrix $\phi$ given in Eq.~(\ref{case1phi}), other choices of eigenvectors also exist which produce the same $\phi$.}
\begin{equation}
\langle \phi \rangle =\left(\begin{matrix}-\frac{1}{2}+\sqrt{2} & 0 & \frac{i}{2}\\
0 & 1 & 0\\
\frac{1}{2} & 0 & -\frac{i}{2}
\end{matrix}\right)
\label{case1phi}
\end{equation}
in matrix form, where the row and the column indices of the matrix correspond to the $(S_4)_1$ and the $(S_4)_2$ indices respectively~. 

From Eq.~(\ref{eq:majorana}) and using Eq.~(\ref{case1phi}) we get the Majorana mass matrix
\begin{align}
M_\text{Maj}&=m_G \langle \phi \rangle\langle \phi \rangle^T \label{eq:majmass}\\
&=m_G\left(\begin{matrix}2-\sqrt{2} & 0 & \frac{1}{\sqrt{2}}\\
0 & 1 & 0\\
\frac{1}{\sqrt{2}} & 0 & 0
\end{matrix}\right)\label{eq:majmass2}
\end{align}
It can be shown that the matrix $\mathcal{T}^\dagger M_\text{Maj}\mathcal{T}^*$, with $M_\text{Maj}$ given in Eq.~(\ref{eq:majmass2}), is a highly constrained form of the complex-symmetric ``Simplest'' texture \cite{Simplest,*S4paper}.

From Eq.~(\ref{eq:dirac}) we get the Dirac mass matrix
\begin{equation}
M_\text{Dir}=m_w I
\label{eq:dirmass}
\end{equation}
where $I$ is the $3\times3$ identity matrix.
If $m_G >> m_w$, then $M_\text{Maj} >> M_\text{Dir}$ resulting in the type-1 see-saw mechanism. The resulting effective see-saw mass matrix \cite{Majorana}, $M_{ss}$, is given by
\begin{equation}
M_{ss}=-M_\text{Dir}M_\text{Maj}^{-1}M_\text{Dir}.
\label{eq:ssmass}
\end{equation}
From Eqs.~(\ref{eq:majmass}-\ref{eq:ssmass}), we get
\begin{align}
M_{ss} &= -k\left(\langle \phi \rangle\langle \phi \rangle^T\right)^{-1} \label{eq:ssmass2}\\
&= -k \left(\begin{matrix}2-\sqrt{2} & 0 & \frac{1}{\sqrt{2}}\\
0 & 1 & 0\\
\frac{1}{\sqrt{2}} & 0 & 0
\end{matrix}\right)^{-1} \label{eq:ssmass3}
\end{align}
where
\begin{equation}
k=\frac{m_w^2}{m_G}.
\end{equation}

The unitary matrix
\begin{equation}
U_\nu=\mathcal{B}_2\mathcal{I}\mathcal{B}_2\mathcal{E}\mathcal{B}_2^T\mathcal{P}
\end{equation}
with
\begin{align}
\mathcal{B}_2&=\left(\begin{matrix}\frac{1}{\sqrt{2}} & 0 & \frac{-1}{\sqrt{2}}\\
0 & 1 & 0\\
\frac{1}{\sqrt{2}} & 0 & \frac{1}{\sqrt{2}}
\end{matrix}\right),\\
\mathcal{I}&=\text{diag}(1,1,i),\\
\mathcal{E}&=\text{diag}(e^{i\frac{\pi}{8}},1,1),\\
\mathcal{P}&=\text{diag}(e^{-i\frac{\pi}{16}},1,e^{-i\frac{\pi}{16}})
\end{align}
diagonalises $M_{ss}$ given in Eq.~(\ref{eq:ssmass3}). In other words we have
\begin{equation}
U_\nu^\dagger M_{ss}U_\nu^*=-\text{diag}\left(m_1,m_2,m_3\right)
\end{equation}
where $m_1$, $m_2$ and $m_3$ are the neutrino masses given by
\begin{align}
m_1&=k\frac{\left(2+\sqrt{2}\right)}{1+\sqrt{2(2+\sqrt{2})}},\notag\\
m_2&=k,\label{numass}\\
m_3&=k\frac{\left(2+\sqrt{2}\right)}{-1+\sqrt{2(2+\sqrt{2})}}.\notag
\end{align}

From $\mathcal{T}$ and $U_\nu$ we obtain the PMNS matrix, U:
\begin{equation}
U=\mathcal{T}U_\nu=\mathcal{T}\mathcal{B}_2\mathcal{I}\mathcal{B}_2\mathcal{E}\mathcal{B}_2^T\mathcal{P}.
\label{eq:pmns1}
\end{equation}
$U$, given in Eq.~(\ref{eq:pmns1}), is a constrained form of the Trichimaximal (T$\chi$M) mixing \cite{TXPM,Xing} with $\chi=\frac{\pi}{16}$:
\begin{equation}
|U|=|\text{T}\chi\text{M}_{\left(\chi=\frac{\pi}{16}\right)}|
\label{eq:txm1}
\end{equation}
where
\begin{equation}
\text{T}\chi\text{M}=\left(\begin{matrix}\sqrt{\frac{2}{3}}\cos \chi & \frac{1}{\sqrt{3}} & \sqrt{\frac{2}{3}}\sin \chi\\
-\frac{\cos \chi}{\sqrt{6}}-i\frac{\sin \chi}{\sqrt{2}} & \frac{1}{\sqrt{3}} & i\frac{\cos \chi}{\sqrt{2}}-\frac{\sin \chi}{\sqrt{6}}\\
-\frac{\cos \chi}{\sqrt{6}}+i\frac{\sin \chi}{\sqrt{2}} & \frac{1}{\sqrt{3}} & -i\frac{\cos \chi}{\sqrt{2}}-\frac{\sin \chi}{\sqrt{6}}
\end{matrix}\right).
\label{eq:txmform}
\end{equation}
The modulus sign, as given in Eq.~(\ref{eq:txm1}), is used throughout this paper to indicate that the expression for the mixing matrix is valid only upto right and left multiplication with diagonal phase matrices (which do not affect the phenomenon of neutrino oscillation). The right multiplying diagonal phase matrices, like $\mathcal{P}$ in Eq.~(\ref{eq:pmns1}), do contribute to Majorana phases, the study of which is beyond the scope of this paper. From Eq.~(\ref{eq:txm1}) and using Eq.~(\ref{eq:txmform}) we get
\begin{gather}
|U_{e3}|^2=\frac{2}{3}\sin^2 \frac{\pi}{16}\,\,\Rightarrow\,\,\sin^2\theta_{13}=0.025,\label{eq:theta13}\\
|U_{e2}|^2=\frac{1}{3}\,\,\Rightarrow\,\,\sin^2\theta_{12}=0.342,\label{eq:theta12}\\
\sin^2\theta_{23}=\frac{1}{2},\label{eq:theta23c1}\\
\delta_{CP}=\frac{\pi}{2}.\label{eq:deltac1}
\end{gather}
From Eqs.~(\ref{numass}) we get the ratios of the neutrino masses,
\begin{equation}
m_1:m_2:m_3=0.945:1:2.117.
\end{equation}
These ratios are consistent with the mass-squared differences measured experimentally \cite{Global1,Global2} within $1\sigma$ errors and thus we can predict the unknown light neutrino mass:
\begin{equation}
24.7~\text{meV}\lesssim m_1\lesssim 25.5~\text{meV}. 
\label{eq:m1}
\end{equation}

Let $x=i^n$ where $n$ is an integer and let 
\begin{equation}
\Phi_{x}=\left(\begin{matrix}i\frac{x}{2}+\frac{1-i x}{\sqrt{2}} & 0 & -\frac{x}{2}\\
0 & 1 & 0\\
-i\frac{x}{2}+\frac{1+i x}{\sqrt{2}} & 0 & \frac{x}{2}
\end{matrix}\right).
\end{equation}
Using $\Phi_x$, Eq.~(\ref{case1phi}) can be rewritten as 
\begin{equation}
\langle \phi \rangle = \Phi_i^*.
\end{equation}
In the following sections we discuss three more cases where $\langle \phi \rangle = \Phi_{-1}^*$, $\Phi_{-i}^*$ and $\Phi_1^*$ respectively. In all these cases, we obtain the same expressions for the neutrino masses as given in Eqs.~(\ref{numass}). The expressions for $\theta_{13}$, Eq.~(\ref{eq:theta13}), and $\theta_{12}$, Eq.~(\ref{eq:theta12}), also remain unchanged. 

\noindent{\it \bf Case 2}: Assigning
\begin{equation}
\begin{aligned}
v^1_1&=e_2, &v^1_2&=e_2,\\
v^2_1&=\text{eig}(c,i)_1, &v^2_2&=e_1,\\
v^3_1&=\text{eig}(c,i)_3, &v^3_2&=e_1,\\
v^4_1&=\text{eig}(d,1)_1, &v^4_2&=\text{eig}(c,\text{-}i)_3
\end{aligned}
\label{eq:case2vev}
\end{equation}
we get 
\begin{align}
\langle \phi \rangle &= \Phi_{i^2}^*\\
&=\left(\begin{matrix}\frac{i}{2}+\frac{1-i}{\sqrt{2}} & 0 & \frac{1}{2}\\
0 & 1 & 0\\
-\frac{i}{2}+\frac{1+i}{\sqrt{2}} & 0 & -\frac{1}{2}
\end{matrix}\right)
\end{align} 
and 
\begin{align}
M_{ss} &= -k\left(\langle \phi \rangle\langle \phi \rangle^T\right)^{-1}\\
&= -k \left(\begin{matrix}-i+\frac{1+i}{\sqrt{2}} & 0 & 1-\frac{1}{\sqrt{2}}\\
0 & 1 & 0\\
1-\frac{1}{\sqrt{2}} & 0 & i+\frac{1-i}{\sqrt{2}}
\end{matrix}\right)^{-1}.
\end{align} 
In this case, we get 
\begin{equation}\label{eq:mix2}
U_\nu=\mathcal{B}_2\mathcal{I}^2\mathcal{B}_2\mathcal{E}\mathcal{B}_2^T\mathcal{P}.
\end{equation}
The resulting PMNS matrix
\begin{equation}\label{eq:mix2}
U=\mathcal{T}U_\nu=\mathcal{T}\mathcal{B}_2\mathcal{I}^2\mathcal{B}_2\mathcal{E}\mathcal{B}_2^T\mathcal{P}
\end{equation}
is a constrained form of the Triphimaximal (T$\phi$M) mixing \cite{TXPM} with $\phi=-\frac{\pi}{16}$:
\begin{equation}
|U|=|\text{T}\phi\text{M}_{\left(\phi=-\frac{\pi}{16}\right)}|.
\label{eq:tpm2}
\end{equation}
where
\begin{equation}
\text{T}\phi\text{M}=\left(\begin{matrix}\sqrt{\frac{2}{3}}\cos \phi & \frac{1}{\sqrt{3}} & \sqrt{\frac{2}{3}}\sin \phi\\
-\frac{\cos \phi}{\sqrt{6}}-\frac{\sin \phi}{\sqrt{2}} & \frac{1}{\sqrt{3}} & \frac{\cos \phi}{\sqrt{2}}-\frac{\sin \phi}{\sqrt{6}}\\
-\frac{\cos \phi}{\sqrt{6}}+\frac{\sin \phi}{\sqrt{2}} & \frac{1}{\sqrt{3}} & -\frac{\cos \phi}{\sqrt{2}}-\frac{\sin \phi}{\sqrt{6}}
\end{matrix}\right).
\label{eq:tpmform}
\end{equation}

From Eq.~(\ref{eq:tpm2}) and using Eq.~(\ref{eq:tpmform}) we get
\begin{gather}
|U_{\mu 3}|^2=\frac{2}{3}\sin^2 \left(\frac{2\pi}{3}-\frac{\pi}{16}\right) \,\, \Rightarrow \,\, \sin^2 \theta_{23}=0.613, \label{eq:theta23c2}\\
\delta_{CP} = \pi. \label{eq:deltac2}
\end{gather}

\noindent{\it \bf Case 3}: The relevant equations are given below:
\begin{equation}
\begin{aligned}
v^1_1&=e_2, &v^1_2&=e_2,\\
v^2_1&=\text{eig}(a,1)_1, &v^2_2&=e_1,\\
v^3_1&=\text{eig}(d,1)_3, &v^3_2&=e_1,\\
v^4_1&=\text{eig}(d,1)_1, &v^4_2&=\text{eig}(c,\text{-}i)_1,
\end{aligned}
\end{equation}
\begin{align}
\langle \phi \rangle &= \Phi_{i^3}^*\\
&=\left(\begin{matrix}\frac{1}{2} & 0 & -\frac{i}{2}\\
0 & 1 & 0\\
-\frac{1}{2}+\sqrt{2} & 0 & \frac{i}{2}
\end{matrix}\right),
\end{align} 
\begin{align}
M_{ss} &= -k\left(\langle \phi \rangle\langle \phi \rangle^T\right)^{-1}\\
&= -k \left(\begin{matrix}0 & 0 & \frac{1}{\sqrt{2}}\\
0 & 1 & 0\\
\frac{1}{\sqrt{2}} & 0 & 2-\sqrt{2}
\end{matrix}\right)^{-1},
\end{align} 
\begin{gather}
U_\nu=\mathcal{B}_2\mathcal{I}^3\mathcal{B}_2\mathcal{E}\mathcal{B}_2^T\mathcal{P},\\
U=\mathcal{T}U_\nu=\mathcal{T}\mathcal{B}_2\mathcal{I}^3\mathcal{B}_2\mathcal{E}\mathcal{B}_2^T\mathcal{P},\label{eq:pmns3}\\
|U|=|\text{T}\chi\text{M}_{\left(\chi=-\frac{\pi}{16}\right)}|,
\end{gather}
\begin{gather}
\sin^2\theta_{23}=\frac{1}{2},\label{eq:theta23c3}\\
\delta_{CP} = \frac{3\pi}{2}.\label{eq:deltac3}
\end{gather}

\noindent{\it \bf Case 4}: The relevant equations are given below:
\begin{equation}
\begin{aligned}
v^1_1&=e_2, &v^1_2&=e_2,\\
v^2_1&=\text{eig}(c,\text{-}i)_1, &v^2_2&=e_1,\\
v^3_1&=\text{eig}(c,\text{-}i)_3, &v^3_2&=e_1,\\
v^4_1&=\text{eig}(d,1)_3, &v^4_2&=\text{eig}(c,\text{-}i)_3,
\end{aligned}
\end{equation}
\begin{align}
\langle \phi \rangle &= \Phi_{i^4}^*\\
&=\left(\begin{matrix}-\frac{i}{2}+\frac{1+i}{\sqrt{2}} & 0 & -\frac{1}{2}\\
0 & 1 & 0\\
\frac{i}{2}+\frac{1-i}{\sqrt{2}} & 0 & \frac{1}{2}
\end{matrix}\right),
\end{align} 
\begin{align}
M_{ss} &= -k\left(\langle \phi \rangle\langle \phi \rangle^T\right)^{-1}\\
&= -k \left(\begin{matrix}i+\frac{1-i}{\sqrt{2}} & 0 & 1-\frac{1}{\sqrt{2}}\\
0 & 1 & 0\\
1-\frac{1}{\sqrt{2}} & 0 & -i+\frac{1+i}{\sqrt{2}}
\end{matrix}\right)^{-1},
\end{align} 
\begin{gather}
U_\nu=\mathcal{B}_2\mathcal{I}^4\mathcal{B}_2\mathcal{E}\mathcal{B}_2^T\mathcal{P},\\
U=\mathcal{T}U_\nu=\mathcal{T}\mathcal{B}_2\mathcal{I}^4\mathcal{B}_2\mathcal{E}\mathcal{B}_2^T\mathcal{P},\label{eq:pmns4}\\
|U|=|\text{T}\phi\text{M}_{\left(\phi=\frac{\pi}{16}\right)}|,
\end{gather}
\begin{gather}
|U_{\mu 3}|^2=\frac{2}{3}\sin^2 \left(\frac{2\pi}{3}+\frac{\pi}{16}\right) \,\, \Rightarrow \,\, \sin^2 \theta_{23}=0.387, \label{eq:theta23c4}\\
\delta_{CP} = 2\pi. \label{eq:deltac4}
\end{gather}

The values predicted by all the four cases of the model are within $3\sigma$ errors of the experimental best fits \cite{Global1,Global2,Global3}. In fact the generic prediction $\sin^2\theta_{13}=0.025$, Eq.~(\ref{eq:theta13}), is within $1\sigma$ errors. However the global analysis \cite{Global1} shows more than $2\sigma$ tension with $\sin^2 \theta_{23}=\frac{1}{2}$, the T$\chi$M value (Cases~1 and 3, Eqs.~(\ref{eq:theta23c1},~\ref{eq:theta23c3})). On the other hand the T$\phi$M values, $\sin^2 \theta_{23}=0.613$ from Eq.~(\ref{eq:theta23c2}) in Case~2 and $\sin^2 \theta_{23}=0.387$ from Eq.~(\ref{eq:theta23c4}) in Case~4, are well within $1\sigma$ errors calculated in \cite{Global2} and \cite{Global1} respectively. All the cases predict $\sin^2\theta_{12}=0.342$, Eq.~(\ref{eq:theta12}), which is at the edge of the $2\sigma$ error range in \cite{Global1}. A new mixing ansatz called the VS mixing~\footnote{Dedicated to my father K.~Venugopal and mother J.~Saraswathi~Amma} is proposed in the following section which modifies $\theta_{12}$ as well as $\delta_{CP}$.
%The global fit \cite{} favours the first octant and the global fits \cite{} favour the second octant for $\theta_{23}$. Hence we modify the mixing by proposing a new ansatz. 

\noindent{\bf The VS Mixing Ansatz}: The mixing obtained using the model, Eqs.~(\ref{eq:pmns1},~\ref{eq:mix2},~\ref{eq:pmns3},~\ref{eq:pmns4}), is of the form
\begin{equation}
|U|=|\mathcal{T}\mathcal{B}_2\mathcal{I}^n\mathcal{B}_2\mathcal{E}\mathcal{B}_2^T|.\label{eq:oldmix}
\end{equation}
The matrix $|\mathcal{T}\mathcal{B}_2\mathcal{I}^n|$ gives the Tribimaximal (TBM) mixing \cite{TBM}. Multiplying $\mathcal{T}\mathcal{B}_2\mathcal{I}^n$ with $\mathcal{B}_2\mathcal{E}\mathcal{B}_2^T$ mixes the first and the third columns of the TBM matrix giving the non-zero value for $\theta_{13}$ in the four cases described in the previous sections. Now we may further mix the first and the second columns of the mixing matrix given Eq.~(\ref{eq:oldmix}). This leaves the last column and as a result $\theta_{13}$ and $\theta_{23}$ unaffected. The resulting new ansatz is defined by
\begin{equation}
|\text{VS}_{i^n}(\alpha)|=|\mathcal{T}\mathcal{B}_2\mathcal{I}^n\mathcal{B}_2\mathcal{E}\mathcal{B}_2^T\mathcal{B}_3\mathcal{E'}\mathcal{B}_3^T|
\label{eq:vsmixing}
\end{equation}
where
\begin{align}
\mathcal{B}_3 &=
\left(\begin{matrix}\frac{1}{\sqrt{2}} & \frac{-1}{\sqrt{2}} & 0\\
       \frac{1}{\sqrt{2}} & \frac{1}{\sqrt{2}} & 0\\
       0 & 0 & 1
\end{matrix}\right)\\
\mathcal{E'}&=\text{diag}(e^{i\alpha},1,1).
\end{align}
Note that the Cases~1 to 4 are simply $\text{VS}_{i^n}(0)$ with $n=1\text{~to~}4$ respectively. Eq.~(\ref{eq:vsmixing}) on simplification gives
\begin{equation}
|\text{VS}_{i^n}(\alpha)|=|\mathcal{T}\mathcal{B}_2\mathcal{I}^n\mathcal{H}_2\mathcal{S}\mathcal{H}'_3|
\end{equation}
where
\begin{gather}
\mathcal{S}=\text{diag}(e^{i\frac{\pi}{16}},1,1),\\
\mathcal{H}_2=\left(\begin{matrix}\boldsymbol{c} & 0 & i \boldsymbol{s}\\
0 & 1 & 0\\
i \boldsymbol{s} & 0 & \boldsymbol{c}
\end{matrix}\right), \quad \mathcal{H}'_3 =
\left(\begin{matrix}\boldsymbol{c}' & i \boldsymbol{s}' & 0\\
       i \boldsymbol{s}' & \boldsymbol{c}' & 0\\
       0 & 0 & 1
\end{matrix}\right)
\end{gather}
with
\begin{equation}
\boldsymbol{c}=\cos\frac{\pi}{16}, \,\,\, \boldsymbol{s}=\sin\frac{\pi}{16}, \,\,\, \boldsymbol{c}'= \cos \frac{\alpha}{2}, \,\,\, \boldsymbol{s}'= \sin \frac{\alpha}{2}.
\end{equation}
We get $\sin^2 \theta_{12}$ within $1\sigma$ errors for $0.08\pi\lesssim\alpha\lesssim0.26\pi$. Table~2 lists a few cases of the VS mixing along with the predicted values of the mixing angles. The author finds the choice $\alpha=\frac{\pi}{8}$ to be aesthetically pleasing. When $\alpha=\frac{\pi}{8}$ we get $\mathcal{E'}=\mathcal{E}$ and also $\boldsymbol{c}'=\boldsymbol{c},~\boldsymbol{s}'=\boldsymbol{s}$.

{\renewcommand{\arraystretch}{1.4}
\begin{table}[H]
\begin{center}
\begin{tabular}{| c | c | c | c | c | c | c |}
\hline
  &$\text{VS}_{\text{-}1}(\tfrac{\pi}{4})$ & $\text{VS}_{1}(\tfrac{\pi}{4})$ & $\text{VS}_{\text{-}1}(\tfrac{\pi}{6})$ & $\text{VS}_{1}(\tfrac{\pi}{6})$ & $\text{VS}_{\text{-}1}(\tfrac{\pi}{8})$ & $\text{VS}_{1}(\tfrac{\pi}{8})$\\
\hline
$\sin^2 \theta_{23}$&$0.613$&$0.387$&$0.613$&$0.387$&$0.613$&$0.387$\\
\hline
$\sin^2 \theta_{12}$&$0.323$&$0.323$&$0.317$&$0.317$&$0.319$&$0.319$\\
\hline
$\delta_{CP}$&$1.27\pi$&$0.27\pi$&$1.18\pi$&$0.18\pi$&$1.13\pi$&$0.13\pi$\\
\hline
\end{tabular}
\end{center}
\caption{Note that $\sin^2 \theta_{13}=0.025$ is a generic feature of the VS mixing. Conjugation, $\text{VS}_{i^n}^*(\alpha)$, changes the sign of $\delta_{CP}$ without affecting the mixing angles $\theta_{12}$, $\theta_{23}$ and $\theta_{13}.$}
\label{tab:VS}
\end{table}}

\noindent{\bf Summary}: The symmetries represented by a discrete group are related to the eigenvectors of the group elements. We develop a notation to uniquely identify the eigenvectors and use it to assign vevs for the flavons. An orthonormal set of eigenvectors define the fermions' flavour states. The model thus constructed predicts the reactor mixing angle, $\sin^2 \theta_{13}=0.025$, and the ratios of the neutrino masses, $m_1:m_2:m_3=0.945:1:2.117$, which are in remarkable agreement with the experimental data. The T$\phi$M versions of the model provide solutions for $\theta_{23}$ in the first octant, $\sin^2\theta_{23}=0.387$, as well as in the second octant, $\sin^2\theta_{23}=0.613$. The T$\phi$M as well as the T$\chi$M versions give $\sin^2\theta_{12}=0.342$. A new mixing ansatz, $\text{VS}_{i^n}(\alpha)$, is introduced which gives reduced values for $\theta_{12}$. The ansatz also predicts various values for $\delta_{CP}$. 

\begin{acknowledgments}
I would like to thank Paul Harrison and Bill Scott for helpful discussions. This work was supported by the UK Science and Technology Facilities Council (STFC). I acknowledge support from the University of Warwick and the Centre for Fundamental Physics at the Rutherford Appleton Laboratory.
\end{acknowledgments}


\begin{thebibliography}{36}%
\makeatletter
\providecommand \@ifxundefined [1]{%
 \@ifx{#1\undefined}
}%
\providecommand \@ifnum [1]{%
 \ifnum #1\expandafter \@firstoftwo
 \else \expandafter \@secondoftwo
 \fi
}%
\providecommand \@ifx [1]{%
 \ifx #1\expandafter \@firstoftwo
 \else \expandafter \@secondoftwo
 \fi
}%
\providecommand \natexlab [1]{#1}%
\providecommand \enquote  [1]{``#1''}%
\providecommand \bibnamefont  [1]{#1}%
\providecommand \bibfnamefont [1]{#1}%
\providecommand \citenamefont [1]{#1}%
\providecommand \href@noop [0]{\@secondoftwo}%
\providecommand \href [0]{\begingroup \@sanitize@url \@href}%
\providecommand \@href[1]{\@@startlink{#1}\@@href}%
\providecommand \@@href[1]{\endgroup#1\@@endlink}%
\providecommand \@sanitize@url [0]{\catcode `\\12\catcode `\$12\catcode
  `\&12\catcode `\#12\catcode `\^12\catcode `\_12\catcode `\%12\relax}%
\providecommand \@@startlink[1]{}%
\providecommand \@@endlink[0]{}%
\providecommand \url  [0]{\begingroup\@sanitize@url \@url }%
\providecommand \@url [1]{\endgroup\@href {#1}{\urlprefix }}%
\providecommand \urlprefix  [0]{URL }%
\providecommand \Eprint [0]{\href }%
\providecommand \doibase [0]{http://dx.doi.org/}%
\providecommand \selectlanguage [0]{\@gobble}%
\providecommand \bibinfo  [0]{\@secondoftwo}%
\providecommand \bibfield  [0]{\@secondoftwo}%
\providecommand \translation [1]{[#1]}%
\providecommand \BibitemOpen [0]{}%
\providecommand \bibitemStop [0]{}%
\providecommand \bibitemNoStop [0]{.\EOS\space}%
\providecommand \EOS [0]{\spacefactor3000\relax}%
\providecommand \BibitemShut  [1]{\csname bibitem#1\endcsname}%
\let\auto@bib@innerbib\@empty
%</preamble>
\bibitem [{\citenamefont {Grimus}\ and\ \citenamefont {Ludl}(2010)}]{SU31}%
  \BibitemOpen
  \bibfield  {author} {\bibinfo {author} {\bibfnamefont {W.}~\bibnamefont
  {Grimus}}\ and\ \bibinfo {author} {\bibfnamefont {P.}~\bibnamefont {Ludl}},\
  }\href@noop {} {\bibfield  {journal} {\bibinfo  {journal} {J. Phys. A}\
  }\textbf {\bibinfo {volume} {43}},\ \bibinfo {pages} {445209} (\bibinfo
  {year} {2010})},\ \Eprint {http://arxiv.org/abs/1006.0098} {arXiv:1006.0098}
  \BibitemShut {NoStop}%
\bibitem [{\citenamefont {King}\ and\ \citenamefont {Ross}(2001)}]{SU32}%
  \BibitemOpen
  \bibfield  {author} {\bibinfo {author} {\bibfnamefont {S.~F.}\ \bibnamefont
  {King}}\ and\ \bibinfo {author} {\bibfnamefont {G.~G.}\ \bibnamefont
  {Ross}},\ }\href@noop {} {\bibfield  {journal} {\bibinfo  {journal} {Phys.
  Lett. B}\ }\textbf {\bibinfo {volume} {520}},\ \bibinfo {pages} {243}
  (\bibinfo {year} {2001})},\ \Eprint {http://arxiv.org/abs/hep-ph/0108112}
  {arXiv:hep-ph/0108112} \BibitemShut {NoStop}%
\bibitem [{\citenamefont {Koide}(2007)}]{SU33}%
  \BibitemOpen
  \bibfield  {author} {\bibinfo {author} {\bibfnamefont {Y.}~\bibnamefont
  {Koide}},\ }\href@noop {} {\  (\bibinfo {year} {2007})},\ \Eprint
  {http://arxiv.org/abs/0707.0899} {arXiv:0707.0899} \BibitemShut {NoStop}%
\bibitem [{\citenamefont {de~Medeiros~Varzielas}\ and\ \citenamefont
  {Ross}(2006)}]{SU34}%
  \BibitemOpen
  \bibfield  {author} {\bibinfo {author} {\bibfnamefont {I.}~\bibnamefont
  {de~Medeiros~Varzielas}}\ and\ \bibinfo {author} {\bibfnamefont {G.~G.}\
  \bibnamefont {Ross}},\ }\href@noop {} {\bibfield  {journal} {\bibinfo
  {journal} {Nucl. Phys. B}\ }\textbf {\bibinfo {volume} {733}},\ \bibinfo
  {pages} {31} (\bibinfo {year} {2006})},\ \Eprint
  {http://arxiv.org/abs/hep-ph/0507176} {arXiv:hep-ph/0507176} \BibitemShut
  {NoStop}%
\bibitem [{\citenamefont {Bazzocchi}\ \emph
  {et~al.}(2009{\natexlab{a}})\citenamefont {Bazzocchi}, \citenamefont
  {Morisi}, \citenamefont {Picariello},\ and\ \citenamefont
  {Torrente-Lujan}}]{SU35}%
  \BibitemOpen
  \bibfield  {author} {\bibinfo {author} {\bibfnamefont {F.}~\bibnamefont
  {Bazzocchi}}, \bibinfo {author} {\bibfnamefont {S.}~\bibnamefont {Morisi}},
  \bibinfo {author} {\bibfnamefont {M.}~\bibnamefont {Picariello}}, \ and\
  \bibinfo {author} {\bibfnamefont {E.}~\bibnamefont {Torrente-Lujan}},\
  }\href@noop {} {\bibfield  {journal} {\bibinfo  {journal} {J. Phys. G}\
  }\textbf {\bibinfo {volume} {36}},\ \bibinfo {pages} {015002} (\bibinfo
  {year} {2009}{\natexlab{a}})},\ \Eprint {http://arxiv.org/abs/0802.1693}
  {arXiv:0802.1693} \BibitemShut {NoStop}%
\bibitem [{\citenamefont {Lam}(2008)}]{S41}%
  \BibitemOpen
  \bibfield  {author} {\bibinfo {author} {\bibfnamefont {C.~S.}\ \bibnamefont
  {Lam}},\ }\href@noop {} {\bibfield  {journal} {\bibinfo  {journal} {Phys.
  Rev. D}\ }\textbf {\bibinfo {volume} {78}},\ \bibinfo {pages} {073015}
  (\bibinfo {year} {2008})},\ \Eprint {http://arxiv.org/abs/0809.1185}
  {arXiv:0809.1185} \BibitemShut {NoStop}%
\bibitem [{\citenamefont {Bazzocchi}\ \emph
  {et~al.}(2009{\natexlab{b}})\citenamefont {Bazzocchi}, \citenamefont
  {Merlo},\ and\ \citenamefont {Morisi}}]{S42}%
  \BibitemOpen
  \bibfield  {author} {\bibinfo {author} {\bibfnamefont {F.}~\bibnamefont
  {Bazzocchi}}, \bibinfo {author} {\bibfnamefont {L.}~\bibnamefont {Merlo}}, \
  and\ \bibinfo {author} {\bibfnamefont {S.}~\bibnamefont {Morisi}},\
  }\href@noop {} {\bibfield  {journal} {\bibinfo  {journal} {Nucl. Phys. B}\
  }\textbf {\bibinfo {volume} {816}},\ \bibinfo {pages} {204} (\bibinfo {year}
  {2009}{\natexlab{b}})},\ \Eprint {http://arxiv.org/abs/0901.2086}
  {arXiv:0901.2086} \BibitemShut {NoStop}%
\bibitem [{\citenamefont {Grimus}\ \emph {et~al.}(2009)\citenamefont {Grimus},
  \citenamefont {Lavoura},\ and\ \citenamefont {Ludl}}]{S43}%
  \BibitemOpen
  \bibfield  {author} {\bibinfo {author} {\bibfnamefont {W.}~\bibnamefont
  {Grimus}}, \bibinfo {author} {\bibfnamefont {L.}~\bibnamefont {Lavoura}}, \
  and\ \bibinfo {author} {\bibfnamefont {P.}~\bibnamefont {Ludl}},\ }\href@noop
  {} {\bibfield  {journal} {\bibinfo  {journal} {J. Phys. G}\ }\textbf
  {\bibinfo {volume} {36}},\ \bibinfo {pages} {115007} (\bibinfo {year}
  {2009})},\ \Eprint {http://arxiv.org/abs/0906.2689} {arXiv:0906.2689}
  \BibitemShut {NoStop}%
\bibitem [{\citenamefont {Yang}\ and\ \citenamefont {Zhang}(2011)}]{S44}%
  \BibitemOpen
  \bibfield  {author} {\bibinfo {author} {\bibfnamefont {R.-Z.}\ \bibnamefont
  {Yang}}\ and\ \bibinfo {author} {\bibfnamefont {H.}~\bibnamefont {Zhang}},\
  }\href@noop {} {\bibfield  {journal} {\bibinfo  {journal} {Phys. Lett. B}\
  }\textbf {\bibinfo {volume} {700}},\ \bibinfo {pages} {316} (\bibinfo {year}
  {2011})},\ \Eprint {http://arxiv.org/abs/1104.0380} {arXiv:1104.0380}
  \BibitemShut {NoStop}%
\bibitem [{\citenamefont {Bazzocchi}\ and\ \citenamefont {Merlo}(2012)}]{S45}%
  \BibitemOpen
  \bibfield  {author} {\bibinfo {author} {\bibfnamefont {F.}~\bibnamefont
  {Bazzocchi}}\ and\ \bibinfo {author} {\bibfnamefont {L.}~\bibnamefont
  {Merlo}},\ }\href@noop {} {\  (\bibinfo {year} {2012})},\ \Eprint
  {http://arxiv.org/abs/1205.5135} {arXiv:1205.5135} \BibitemShut {NoStop}%
\bibitem [{\citenamefont {Coxeter}\ and\ \citenamefont {Moser}(1972)}]{book}%
  \BibitemOpen
  \bibfield  {author} {\bibinfo {author} {\bibfnamefont {H.~S.~M.}\
  \bibnamefont {Coxeter}}\ and\ \bibinfo {author} {\bibfnamefont {W.~O.~J.}\
  \bibnamefont {Moser}},\ }\href@noop {} {\emph {\bibinfo {title} {Generators
  and Relations for Discrete Groups}}}\ (\bibinfo  {publisher}
  {Springer-Verlag},\ \bibinfo {year} {1972})\BibitemShut {NoStop}%
\bibitem [{\citenamefont {{\it et al.}~(Daya
  Bay~Collaboration)}(2012)}]{DayaBay}%
  \BibitemOpen
  \bibfield  {author} {\bibinfo {author} {\bibfnamefont {F.~P.~An}\
  \bibnamefont {{\it et al.}~(Daya Bay~Collaboration)}},\ }\href@noop {}
  {\bibfield  {journal} {\bibinfo  {journal} {Phys. Rev. Lett.}\ }\textbf
  {\bibinfo {volume} {108}},\ \bibinfo {pages} {171803} (\bibinfo {year}
  {2012})},\ \Eprint {http://arxiv.org/abs/1203.1669} {arXiv:1203.1669}
  \BibitemShut {NoStop}%
\bibitem [{\citenamefont {{\it et al.}~(RENO~Collaboration)}(2012)}]{RENO}%
  \BibitemOpen
  \bibfield  {author} {\bibinfo {author} {\bibfnamefont {J.~K.~Ahn}\
  \bibnamefont {{\it et al.}~(RENO~Collaboration)}},\ }\href@noop {} {\bibfield
   {journal} {\bibinfo  {journal} {Phys. Rev. Lett.}\ }\textbf {\bibinfo
  {volume} {108}},\ \bibinfo {pages} {191802} (\bibinfo {year} {2012})},\
  \Eprint {http://arxiv.org/abs/1204.0626} {arXiv:1204.0626} \BibitemShut
  {NoStop}%
\bibitem [{\citenamefont {{\it et al.}~(Double
  Chooz~Collaboration)}(2012)}]{DCHOOZ}%
  \BibitemOpen
  \bibfield  {author} {\bibinfo {author} {\bibfnamefont {Y.~Abe}\ \bibnamefont
  {{\it et al.}~(Double Chooz~Collaboration)}},\ }\href@noop {} {\bibfield
  {journal} {\bibinfo  {journal} {Phys. Rev. Lett.}\ }\textbf {\bibinfo
  {volume} {108}},\ \bibinfo {pages} {131801} (\bibinfo {year} {2012})},\
  \Eprint {http://arxiv.org/abs/1112.6353} {arXiv:1112.6353} \BibitemShut
  {NoStop}%
\bibitem [{\citenamefont {{\it et al.}~(T2K~Collaboration)}(2011)}]{T2K}%
  \BibitemOpen
  \bibfield  {author} {\bibinfo {author} {\bibfnamefont {K.~Abe}\ \bibnamefont
  {{\it et al.}~(T2K~Collaboration)}},\ }\href@noop {} {\bibfield  {journal}
  {\bibinfo  {journal} {Phys. Rev. Lett.}\ }\textbf {\bibinfo {volume} {107}},\
  \bibinfo {pages} {041801} (\bibinfo {year} {2011})},\ \Eprint
  {http://arxiv.org/abs/1106.2822} {arXiv:1106.2822} \BibitemShut {NoStop}%
\bibitem [{\citenamefont {{\it et al.}~(MINOS~Collaboration)}(2011)}]{MINOS}%
  \BibitemOpen
  \bibfield  {author} {\bibinfo {author} {\bibfnamefont {P.~Adamson}\ \bibnamefont
  {{\it et al.}~(MINOS~Collaboration)}},\ }\href@noop {} {\bibfield  {journal}
  {\bibinfo  {journal} {Phys. Rev. Lett.}\ }\textbf {\bibinfo {volume} {107}},\
  \bibinfo {pages} {181802} (\bibinfo {year} {2011})},\ \Eprint
  {http://arxiv.org/abs/1108.0015} {arXiv:1108.0015} \BibitemShut {NoStop}%
\bibitem [{\citenamefont {Altarelli}\ \emph {et~al.}(2012)\citenamefont
  {Altarelli}, \citenamefont {Feruglio}, \citenamefont {Merlo},\ and\
  \citenamefont {Stamou}}]{M0}%
  \BibitemOpen
  \bibfield  {author} {\bibinfo {author} {\bibfnamefont {G.}~\bibnamefont
  {Altarelli}}, \bibinfo {author} {\bibfnamefont {F.}~\bibnamefont {Feruglio}},
  \bibinfo {author} {\bibfnamefont {L.}~\bibnamefont {Merlo}}, \ and\ \bibinfo
  {author} {\bibfnamefont {E.}~\bibnamefont {Stamou}},\ }\href@noop {}
  {\bibfield  {journal} {\bibinfo  {journal} {JHEP}\ }\textbf {\bibinfo
  {volume} {1208}},\ \bibinfo {pages} {021} (\bibinfo {year} {2012})},\ \Eprint
  {http://arxiv.org/abs/1205.4670} {arXiv:1205.4670} \BibitemShut {NoStop}%
\bibitem [{\citenamefont {Rodejohann}\ and\ \citenamefont {Zhang}(2012)}]{M1}%
  \BibitemOpen
  \bibfield  {author} {\bibinfo {author} {\bibfnamefont {W.}~\bibnamefont
  {Rodejohann}}\ and\ \bibinfo {author} {\bibfnamefont {H.}~\bibnamefont
  {Zhang}},\ }\href@noop {} {\bibfield  {journal} {\bibinfo  {journal} {Phys.
  Rev. D}\ }\textbf {\bibinfo {volume} {86}},\ \bibinfo {pages} {093008}
  (\bibinfo {year} {2012})},\ \Eprint {http://arxiv.org/abs/1207.1225}
  {arXiv:1207.1225} \BibitemShut {NoStop}%
\bibitem [{\citenamefont {Patel}(2011)}]{M2}%
  \BibitemOpen
  \bibfield  {author} {\bibinfo {author} {\bibfnamefont {K.~M.}\ \bibnamefont
  {Patel}},\ }\href@noop {} {\bibfield  {journal} {\bibinfo  {journal} {Phys.
  Lett. B}\ }\textbf {\bibinfo {volume} {695}},\ \bibinfo {pages} {225}
  (\bibinfo {year} {2011})},\ \Eprint {http://arxiv.org/abs/1008.5061}
  {arXiv:1008.5061} \BibitemShut {NoStop}%
\bibitem [{\citenamefont {Meloni}(2011)}]{M3}%
  \BibitemOpen
  \bibfield  {author} {\bibinfo {author} {\bibfnamefont {D.}~\bibnamefont
  {Meloni}},\ }\href@noop {} {\bibfield  {journal} {\bibinfo  {journal} {JHEP}\
  }\textbf {\bibinfo {volume} {10}},\ \bibinfo {pages} {010} (\bibinfo {year}
  {2011})},\ \Eprint {http://arxiv.org/abs/1107.0221} {arXiv:1107.0221}
  \BibitemShut {NoStop}%
\bibitem [{\citenamefont {King}\ \emph {et~al.}(2012)\citenamefont {King},
  \citenamefont {Luhn},\ and\ \citenamefont {Stuart}}]{M4}%
  \BibitemOpen
  \bibfield  {author} {\bibinfo {author} {\bibfnamefont {S.~F.}\ \bibnamefont
  {King}}, \bibinfo {author} {\bibfnamefont {C.}~\bibnamefont {Luhn}}, \ and\
  \bibinfo {author} {\bibfnamefont {A.~J.}\ \bibnamefont {Stuart}},\
  }\href@noop {} {\  (\bibinfo {year} {2012})},\ \Eprint
  {http://arxiv.org/abs/1207.5741} {arXiv:1207.5741} \BibitemShut {NoStop}%
\bibitem [{\citenamefont {Lin}(2010)}]{M5}%
  \BibitemOpen
  \bibfield  {author} {\bibinfo {author} {\bibfnamefont {Y.}~\bibnamefont
  {Lin}},\ }\href@noop {} {\bibfield  {journal} {\bibinfo  {journal} {Nucl.
  Phys. B}\ }\textbf {\bibinfo {volume} {824}},\ \bibinfo {pages} {95}
  (\bibinfo {year} {2010})},\ \Eprint {http://arxiv.org/abs/0905.3534}
  {arXiv:0905.3534} \BibitemShut {NoStop}%
\bibitem [{\citenamefont {Morisi}\ \emph {et~al.}(2011)\citenamefont {Morisi},
  \citenamefont {Patel},\ and\ \citenamefont {Peinado}}]{M6}%
  \BibitemOpen
  \bibfield  {author} {\bibinfo {author} {\bibfnamefont {S.}~\bibnamefont
  {Morisi}}, \bibinfo {author} {\bibfnamefont {K.~M.}\ \bibnamefont {Patel}}, \
  and\ \bibinfo {author} {\bibfnamefont {E.}~\bibnamefont {Peinado}},\
  }\href@noop {} {\bibfield  {journal} {\bibinfo  {journal} {Phys. Rev. D}\
  }\textbf {\bibinfo {volume} {84}},\ \bibinfo {pages} {053002} (\bibinfo
  {year} {2011})},\ \Eprint {http://arxiv.org/abs/1107.0696} {arXiv:1107.0696}
  \BibitemShut {NoStop}%
\bibitem [{\citenamefont {Ishimori}\ \emph {et~al.}(2010)\citenamefont
  {Ishimori}, \citenamefont {Kobayashi}, \citenamefont {Ohki}, \citenamefont
  {Shimizu}, \citenamefont {Okada},\ and\ \citenamefont {Tanimoto}}]{S4Table}%
  \BibitemOpen
  \bibfield  {author} {\bibinfo {author} {\bibfnamefont {H.}~\bibnamefont
  {Ishimori}}, \bibinfo {author} {\bibfnamefont {T.}~\bibnamefont {Kobayashi}},
  \bibinfo {author} {\bibfnamefont {H.}~\bibnamefont {Ohki}}, \bibinfo {author}
  {\bibfnamefont {Y.}~\bibnamefont {Shimizu}}, \bibinfo {author} {\bibfnamefont
  {H.}~\bibnamefont {Okada}}, \ and\ \bibinfo {author} {\bibfnamefont
  {M.}~\bibnamefont {Tanimoto}},\ }\href@noop {} {\bibfield  {journal}
  {\bibinfo  {journal} {Prog. Theor. Phys. Suppl.}\ }\textbf {\bibinfo {volume}
  {183}},\ \bibinfo {pages} {1} (\bibinfo {year} {2010})},\ \Eprint
  {http://arxiv.org/abs/1003.3552} {arXiv:1003.3552} \BibitemShut {NoStop}%
\bibitem [{\citenamefont {Harrison}\ \emph {et~al.}(1999)\citenamefont
  {Harrison}, \citenamefont {Perkins},\ and\ \citenamefont {Scott}}]{TM}%
  \BibitemOpen
  \bibfield  {author} {\bibinfo {author} {\bibfnamefont {P.~F.}\ \bibnamefont
  {Harrison}}, \bibinfo {author} {\bibfnamefont {D.~H.}\ \bibnamefont
  {Perkins}}, \ and\ \bibinfo {author} {\bibfnamefont {W.~G.}\ \bibnamefont
  {Scott}},\ }\href@noop {} {\bibfield  {journal} {\bibinfo  {journal} {Phys.
  Lett. B}\ }\textbf {\bibinfo {volume} {458}},\ \bibinfo {pages} {79}
  (\bibinfo {year} {1999})},\ \Eprint {http://arxiv.org/abs/hep-ph/9904297}
  {arXiv:hep-ph/9904297} \BibitemShut {NoStop}%
\bibitem [{Note1()}]{Note1}%
  \BibitemOpen
  \bibinfo {note} {The set of eigenvectors given in Eqs.~(\ref {eq:case1vev})
  is not the only choice that results in the matrix $\langle\phi\rangle$ given in Eq.~(\ref
  {case1phi}), other choices of eigenvectors also exist which produce the same
  $\langle\phi\rangle$.}\BibitemShut {Stop}%
\bibitem [{\citenamefont {Harrison}\ and\ \citenamefont
  {Scott}(2004)}]{Simplest}%
  \BibitemOpen
  \bibfield  {author} {\bibinfo {author} {\bibfnamefont {P.~F.}\ \bibnamefont
  {Harrison}}\ and\ \bibinfo {author} {\bibfnamefont {W.~G.}\ \bibnamefont
  {Scott}},\ }\href@noop {} {\bibfield  {journal} {\bibinfo  {journal} {Phys.
  Lett. B}\ }\textbf {\bibinfo {volume} {594}},\ \bibinfo {pages} {324}
  (\bibinfo {year} {2004})},\ \Eprint {http://arxiv.org/abs/hep-ph/0403278}
  {arXiv:hep-ph/0403278} \BibitemShut {NoStop}%
\bibitem [{\citenamefont {Krishnan}\ \emph {et~al.}(2012)\citenamefont
  {Krishnan}, \citenamefont {Harrison},\ and\ \citenamefont {Scott}}]{S4paper}%
  \BibitemOpen
  \bibfield  {author} {\bibinfo {author} {\bibfnamefont {R.}~\bibnamefont
  {Krishnan}}, \bibinfo {author} {\bibfnamefont {P.~F.}\ \bibnamefont
  {Harrison}}, \ and\ \bibinfo {author} {\bibfnamefont {W.~G.}\ \bibnamefont
  {Scott}},\ }\href@noop {} {\  (\bibinfo {year} {2012})},\ \Eprint
  {http://arxiv.org/abs/1211.2000} {arXiv:1211.2000} \BibitemShut {NoStop}%
\bibitem [{\citenamefont {Smirnov}(1993)}]{Majorana}%
  \BibitemOpen
  \bibfield  {author} {\bibinfo {author} {\bibfnamefont {A.~Y.}\ \bibnamefont
  {Smirnov}},\ }\href@noop {} {\bibfield  {journal} {\bibinfo  {journal} {Phys.
  Rev. D}\ }\textbf {\bibinfo {volume} {48}},\ \bibinfo {pages} {3264}
  (\bibinfo {year} {1993})},\ \Eprint {http://arxiv.org/abs/hep-ph/9304205}
  {arXiv:hep-ph/9304205} \BibitemShut {NoStop}%
\bibitem [{\citenamefont {Harrison}\ and\ \citenamefont {Scott}(2002)}]{TXPM}%
  \BibitemOpen
  \bibfield  {author} {\bibinfo {author} {\bibfnamefont {P.~F.}\ \bibnamefont
  {Harrison}}\ and\ \bibinfo {author} {\bibfnamefont {W.~G.}\ \bibnamefont
  {Scott}},\ }\href@noop {} {\bibfield  {journal} {\bibinfo  {journal} {Phys.
  Lett. B}\ }\textbf {\bibinfo {volume} {535}},\ \bibinfo {pages} {163}
  (\bibinfo {year} {2002})},\ \Eprint {http://arxiv.org/abs/hep-ph/0203209}
  {arXiv:hep-ph/0203209} \BibitemShut {NoStop}%
\bibitem [{\citenamefont {zhong Xing}(2002)}]{Xing}%
  \BibitemOpen
  \bibfield  {author} {\bibinfo {author} {\bibfnamefont {Z.}~\bibnamefont
  {zhong Xing}},\ }\href@noop {} {\bibfield  {journal} {\bibinfo  {journal}
  {Phys. Lett. B}\ }\textbf {\bibinfo {volume} {533}},\ \bibinfo {pages} {85}
  (\bibinfo {year} {2002})},\ \Eprint {http://arxiv.org/abs/hep-ph/0204049}
  {arXiv:hep-ph/0204049} \BibitemShut {NoStop}%
\bibitem [{\citenamefont {Fogli}\ \emph {et~al.}(2012)\citenamefont {Fogli},
  \citenamefont {Lisi}, \citenamefont {Marrone}, \citenamefont {Montanino},
  \citenamefont {Palazzo},\ and\ \citenamefont {Rotunno}}]{Global1}%
  \BibitemOpen
  \bibfield  {author} {\bibinfo {author} {\bibfnamefont {G.~L.}\ \bibnamefont
  {Fogli}}, \bibinfo {author} {\bibfnamefont {E.}~\bibnamefont {Lisi}},
  \bibinfo {author} {\bibfnamefont {A.}~\bibnamefont {Marrone}}, \bibinfo
  {author} {\bibfnamefont {D.}~\bibnamefont {Montanino}}, \bibinfo {author}
  {\bibfnamefont {A.}~\bibnamefont {Palazzo}}, \ and\ \bibinfo {author}
  {\bibfnamefont {A.~M.}\ \bibnamefont {Rotunno}},\ }\href@noop {} {\bibfield
  {journal} {\bibinfo  {journal} {Phys. Rev. D}\ }\textbf {\bibinfo {volume}
  {86}},\ \bibinfo {pages} {013012} (\bibinfo {year} {2012})},\ \Eprint
  {http://arxiv.org/abs/1205.5254} {arXiv:1205.5254} \BibitemShut {NoStop}%
\bibitem [{\citenamefont {Forero}\ \emph {et~al.}(2012)\citenamefont {Forero},
  \citenamefont {Tortola},\ and\ \citenamefont {Valle}}]{Global2}%
  \BibitemOpen
  \bibfield  {author} {\bibinfo {author} {\bibfnamefont {D.~V.}\ \bibnamefont
  {Forero}}, \bibinfo {author} {\bibfnamefont {M.}~\bibnamefont {Tortola}}, \
  and\ \bibinfo {author} {\bibfnamefont {J.~W.~F.}\ \bibnamefont {Valle}},\
  }\href@noop {} {\bibfield  {journal} {\bibinfo  {journal} {Phys. Rev. D}\
  }\textbf {\bibinfo {volume} {86}},\ \bibinfo {pages} {073012} (\bibinfo
  {year} {2012})},\ \Eprint {http://arxiv.org/abs/1205.4018} {arXiv:1205.4018}
  \BibitemShut {NoStop}%
\bibitem [{\citenamefont {Gonzalez-Garcia}\ \emph {et~al.}(2012)\citenamefont
  {Gonzalez-Garcia}, \citenamefont {Maltoni}, \citenamefont {Salvado},\ and\
  \citenamefont {Schwetz}}]{Global3}%
  \BibitemOpen
  \bibfield  {author} {\bibinfo {author} {\bibfnamefont {M.~C.}\ \bibnamefont
  {Gonzalez-Garcia}}, \bibinfo {author} {\bibfnamefont {M.}~\bibnamefont
  {Maltoni}}, \bibinfo {author} {\bibfnamefont {J.}~\bibnamefont {Salvado}}, \
  and\ \bibinfo {author} {\bibfnamefont {T.}~\bibnamefont {Schwetz}},\
  }\href@noop {} {\  (\bibinfo {year} {2012})},\ \Eprint
  {http://arxiv.org/abs/1209.3023} {arXiv:1209.3023} \BibitemShut {NoStop}%
\bibitem [{Note2()}]{Note2}%
  \BibitemOpen
  \bibinfo {note} {Dedicated to my father K.~Venugopal and mother
  J.~Saraswathi~Amma}\BibitemShut {NoStop}%
\bibitem [{\citenamefont {Harrison}\ \emph {et~al.}(2002)\citenamefont
  {Harrison}, \citenamefont {Perkins},\ and\ \citenamefont {Scott}}]{TBM}%
  \BibitemOpen
  \bibfield  {author} {\bibinfo {author} {\bibfnamefont {P.~F.}\ \bibnamefont
  {Harrison}}, \bibinfo {author} {\bibfnamefont {D.~H.}\ \bibnamefont
  {Perkins}}, \ and\ \bibinfo {author} {\bibfnamefont {W.~G.}\ \bibnamefont
  {Scott}},\ }\href@noop {} {\bibfield  {journal} {\bibinfo  {journal} {Phys.
  Lett. B}\ }\textbf {\bibinfo {volume} {530}},\ \bibinfo {pages} {167}
  (\bibinfo {year} {2002})},\ \Eprint {http://arxiv.org/abs/hep-ph/0202074}
  {arXiv:hep-ph/0202074} \BibitemShut {NoStop}%
\end{thebibliography}
\end{document}